# Bulk Polycrystalline Ceria Doped Al$_2$O$_3$ and YAG Ceramics for High-Power Density Laser-Driven Solid-State White Lighting: Effects of Crystallinity and Extreme Temperatures

E. H. Penilla*, P. Sellappan†, M.A. Duarte, A.T. Wieg, M. Wingert, and J. E. Garay

Advanced Materials Processing and Synthesis (AMPS) Laboratory, Materials Science & Engineering and Department of Mechanical and Aerospace Engineering, University of California, San Diego, California 92093 USA

*Corresponding author: eh.penilla@gmail.com,
† - Currently at Rolls-Royce High Temperature Composites, Cypress, CA

**ABSTRACT:**

Here we develop and characterize high thermal conductivity/high thermal shock resistant bulk Ce doped Al$_2$O$_3$ and propose it as a new phosphor converting capping layer for high-powered/high-brightness solid-state white lighting (SSWL). The bulk, dense Ce:Al$_2$O$_3$ ceramics have a 0.5 at.% Ce:Al concentration (significantly higher than the equilibrium solubility limit), and were produced using a simultaneous solid-state reactive Current Activated Pressure-Assisted Densification (CAPAD) approach. Ce:Al$_2$O$_3$ exhibits a broadband emission from 400-600nm, which encompasses the entire blue and green portions of the visible spectrum when pumped with ultra-violet (UV) light that is now commercially available in UV light emitting devices (LED) and laser diodes (LD). These broadband phosphors can be used in the commonly employed scheme of mixing with other UV converting capping layers that emit red light to produce white light. Alternatively, they can be used in a novel composite down converter approach that ensures improved thermal-mechanical properties of the converting phosphor capping layer. In this configuration Ce:Al$_2$O$_3$ is used with proven phosphor conversion materials such as Ce:YAG as an active encapsulant or as a capping layer to produce SSWL with an improved bandwidth in the blue portion of the visible spectrum. In order to study the effect of crystallinity on the Ce PL emission, we synthesize Ce:YAG ceramics using high-pressure CAPAD at moderate temperatures to obtain varying crystallinity (amorphous through fully -crystalline). We investigate the PL characteristics of Ce:Al$_2$O$_3$ and Ce:YAG from 295K to 4K, revealing unique crystal field effects from the matrix on the Ce-dopants. The unique PL properties in conjunction with the superior thermal-mechanical properties of Ce:Al$_2$O$_3$ can be used in high-powered/high-brightness integrated devices based on high-efficiency UV-LD that do not suffer from efficiency droop at high drive currents to pump the solid-state capping phosphors.

**I. INTRODUCTION**

The development of high-performance solid-state white lighting (SSWL) materials and devices remains intense, with phosphor down-converted LED's a dominant technology. The approach of using InGaN-based blue LED down-converted with Ce:YAG has proven extremely successful commercially [1]. In order to meet the ever-increasing needs for higher-brightness and high-efficiency integrated devices, designs are trending toward using semi-conductor laser diodes (LD) as a pumping source because high-pump powers lead to increased down-converted brightness without sacrifices in efficiency since unlike LEDs, LD do not suffer from efficiency droop at high drive currents[2].

The higher power promised by LD white light however leads to significant thermal management concerns at the integrated device level that are often dictated by the thermal properties of the phosphor capping layers. The higher pump powers and power densities lead to localized heating stemming from quantum defect during down-conversion, which can degrade the photoluminescent (PL) properties through temperature dependent quenching effects for example[3]. In addition, aggressive pumping ultimately leads to thermal gradients within the phosphor capping layer(s) that may result in melting and/or thermal fracture. Melting of the capping layer is particularly prevalent in low thermal conductivity materials such as polymer-phosphor composites which have k<1 W/(m·K), while thermal fracture can occur in bulk glasses and ceramics which have moderate thermal conductivities (1-2 glasses[4], and ~10 W/(m·K) for RE:YAG[5])

In order to solve these thermal managements problems, SSWL-device engineers are relying upon strategies developed to address similar thermal management problems observed in monochromatic light conversion sources, i.e. high-powered solid-state lasers. These methods include 1) using conversion phosphor materials with improved thermomechanical properties and 2) relying upon more sophisticated cooling schemes to reduce the localized thermal load. This is because the deliverable power of the converter scales with the thermal conductivity $k$, and the fracture stress $\sigma_F$, with the thermal shock figure of merit given by:

$$R_S = \frac{k(1-v)}{\alpha E}\sigma_F, \qquad \text{(Eq. 1)}$$

where E is the elastic modulus, α is the coefficient of thermal expansion and $v$ is Poisson's ratio.

The materials issues are being addressed by 1) shifting away from embedding active phosphors like Ce:YAG within polymer epoxy and using instead glasses as encapsulants[6], 2) using bulk ceramic or single crystal Ce:YAG[7] and 3) using high-thermally conductivity passive ceramics as a phosphor encapsulant, i.e. an $Al_2O_3$/Ce:YAG composite[8]–[11], which result in a higher composite thermal conductivity. The management can be addressed by using different heat sink materials for the LED packaging[12]. In addition, many different hosts for Ce have been investigated as suitable candidates to replace YAG, such as the glasses[13], other garnets[14] and sesquioxides[15] because their relatively open structure permits high solubility (~5%) of rare-earth ions [16]. Many other rare-earths and transition metals have also been doped into many crystal hosts to produce phosphors for down-converting blue and UV pumping[17].

However, with few exceptions[18], it has proven difficult to find another single material system that possesses such a broad PL spectrum as Ce:YAG. The different crystal field interactions result in doped phosphors with unique optical properties. This leads to designs reliant on mixtures of phosphors, each with PL in different portions of the visible spectrum (red, green, blue, RGB) being stacked or mixed to produce down-converted SSWL. The differing PL efficiencies, quantum defect, and sensitivities to temperature can result in poor color rendering, effects which are only amplified under aggressive pumping, i.e. thermal loading[1][12].

We propose the use of polycrystalline sapphire (alumina), as a host for the Ce atom, because of its superior thermomechanical properties to most host materials will enable similar aggressive optical pumping of LD-pumped broadband phosphors. Alumina is a well-known PL host, especially for transition metal dopants Cr and Ti, which are commercially known as synthetic ruby and Ti:sapphire, arguably the most important gain media[19]. Ruby was used to first demonstrate the laser[20] and Ti:sapphire[21] is still used to produce the shortest temporal optical laser pulses. The crystal field interactions of alumina with each dopant results in starkly different PL properties. Ruby is characterized by dominant monochromatic transitions amenable to high-energy lasing with extremely narrow line-width (5Å) and Ti:sapphire possess an extremely large gain bandwidth (520nm), which is useful for producing tunable laser sources and/or ultra-short pulses.

The ability to dope Ce into alumina could result in broadband phosphor with properties suitable for high-power/high-brightness SSWL devices. The challenge in achieving such a material is with incorporating adequate quantities of the Ce into the alumina matrix to sufficiently down-convert the pumping source. This is because of the ionic radii disparity of rare-earths with $Al^{3+}$ (dopant host site) leads to an extremely low-equilibrium solubility limit of ~$10^{-3}$% for the rare-earths far below the concentrations used for down-conversion phosphors.[22] Stemming from our earlier advancements in producing bulk ceramic Ce:YAG[23] and $Tb^{3+}$[24] and $Nd^{3+}$[25] doped ceramic alumina for high-energy/high-powered solid-state laser gain media, we have produced highly-doped bulk polycrystalline $Ce:Al_2O_3$, using similar simultaneous solid-state reactive-CAPAD processing that permits access to kinetic windows away from equilibrium conditions. This processing has permitted the doping of 0.5at.% (Ce:Al), a concentration about 500 times above the equilibrium solubility limit for rare-earths in alumina.

We show that the unique crystal field results in a broadband emission from 400-600nm from the $Ce^{3+}$ dopant, when it is pumped with ultra-violet (UV) light at 350nm, allowing its use as a down-converter. Conveniently, this pumping wavelength is now widely commercially available in UV LED/LD. Conceptually this is shown in **Figure 1a** while the PL under UV excitation at 355nm from a 6ns laser pumping pulse is shown in **Figure 1b**. The broadband emission that covers the blue, green, (notice the blue green halo) and extents into the yellow portion of the visible spectrum results in a near white hue although it does lack sufficient red contribution required for a true white. To increase red components, $Ce:Al_2O_3$ can be used in conjunction with other capping layers that produce red light to produce white light via color mixing and/or with traditional capping layer materials such as Ce:YAG to produce an integrated device with an improved bandwidth in the blue portion of the visible spectrum. This can be accomplished as a layered stack (**Figure 1c**, with resultant PL in **Figure 1d**) or as a homogenously mixed $Al_2O_3$-Ce:YAG/ceramic (**Figure 1e**) composite similar to that demonstrated by Denault[9], Cozzan[8] and Wang[11] to make a bulk phosphor with improved thermomechanical properties. In this proposed scenario, the alumina portion would not only serve as a high-thermal conductivity encapsulant, but also as an active phosphor within the bulk ceramic. The overlap in the respective PL emission line-shapes from $Ce:Al_2O_3$ and Ce:YAG can be appreciated in **Figure 1f** (details in **Section III-D**) and should result in similar white light as from the layered structure shown in **Figure 1d.**

In order to understand the crystal field effects in $Ce:Al_2O_3$, we have characterized the PL response from 295K to 4K, revealing the effects of thermal broadening in this crystal system. We have also synthesized Ce:YAG ceramics powders through chemical precursor approach and processed bulk, dense materials using high-pressure, Current Activated Pressure Assisted Densification (CAPAD) route at moderate temperatures to obtain variance in bulk crystallinity, *i.e.* unique crystal fields. Previous studies by V. Jayaram et al., in various $Al_2O_3$-$Y_2O_3$ systems has shown that it is possible to produce bulk, dense, amorphous, partially crystalline and nano crystalline YAG possible under similar processing temperature and pressure conditions[26][27].

Here we study the effects of tailored crystallinity on the PL properties of bulk Ce:YAG ceramics. We also have investigated the PL response of these samples in a similar temperature range. The PL characteristics of Ce:Al$_2$O$_3$ and Ce:YAG show opposite trends in PL intensity response as a function of sample temperature; the emission intensity of Ce:YAG decreases while Ce:Al$_2$O$_3$ increases as temperature decreases. The PL characterization allows us to propose the band diagram for Ce:Al$_2$O$_3$ and the comparison to Ce:YAG indicates that the 5d levels in Ce:Al2O3 likely do not overlap with the conduction band as in the case for Ce:YAG. The increase in the PL intensity at lower temperatures for Ce:Al$_2$O$_3$ is indicative of thermal broadening effects that are observed in rare-earth doped oxide crystals and glasses. The combined thermomechanical and PL properties of Ce:Al$_2$O$_3$ may permit its use in high-powered/high-brightness integrated devices that use high-efficiency LD as efficient pumping sources for the down-conversion phosphors.

## II. Results and Discussion
### A. Phase, Microstructural, And Optical Analysis of Ce:YAG Powders

TGA/DTA results showed loss of volatile matters present in the as synthesized powders results significant weight reduction, ~60wt.%, when the powders heat treated at the rate of 30°C/min over 500°C (**Figure 2a**). After 700°C, the rate of loss is minimal and with the majority of volatile materials removed. Powder X-ray analysis (**Figure 2b**) indicates that the Ce:YAG powders prepared by organic/inorganic steric entrapment (the PVA synthesis route) had low crystallinity (remained X-ray amorphous), even after calcining at 800°C for 30 minutes. XRD analysis also indicates that the absence of the common perovskite phase, YAP, which is commonly observed during the (pristine or doped) YAG synthesis[28]. The direct formation of garnet structure from amorphous powder behavior was also observed in co-precipitation of pristine YAG powders[26]. Upon further heat treatment above 850°C, the amorphous powders starts crystallize and completely converted into the desired garnet structure (Ce:YAG) and no remanence of perovskite structure, Y$_2$O$_3$ or Al$_2$O$_3$ can be observed.

In the case of Ce:YAG, synthesized through chemical precursor route offers excellent mixing of the starting precursors and high phase purity of the resultant oxides[29]. In addition, with low processing temperature compared to conventional routes, it also results materials with highly porous in nature[30]. The microstructure of the powder can be seen in SEM micrographs of powders calcined at 700°C for 30min shown in **Figure 2c and 2d**. The fine nano-porous microstructure of the powder is typical of the steric entrapment synthesis method which facilitates the grinding/milling process[30][31]. The finely grounded/milled powders tends to densify due to particle sliding and re-arrangement under relatively higher pressure at moderate temperature regimes which contribute to complete densification of materials without crystallization or partial crystallization[26][27][31].

**B. Phase and Microstructural Analysis of Bulk Ce:YAG Ceramics**

Materials with varying degrees of crystallinity are useful for investigating crystal-field effects associated with bulk consolidation. To this aim, the as synthesized Ce:YAG powders were consolidated with CAPAD at varying temperatures as described in **Section IV-B**. **Figure 2a** shows that the powders densified at 500°C for 15 minutes under 500MPa shows very low crystallinity. When the consolidation temperature increased to 700°C, partial crystallization is observed as evidenced by the presence of YAG phase along with remnant amorphous powders. XRD pattern of the partially crystallized peaks indicated, crystallization of pure YAG. As previously reported for the co-precipitated pristine YAG powders[26], in the current study, Ce doped YAG powders were also consolidated under high pressure at moderately low temperature. In order to produce a fully dense and also completely crystallized material, the consolidation temperature was increased to 1400°C, under 100 MPa. Reduction of pressure from 500 to 100MPa is necessary to avoid deformation of the die materials. Bulk densities values measured through Archimedes method for all the bulk samples are plotted against the processing temperature in **Figure 3b**. Fully crystallized material (1400°C/100 MPa) is close to 99.1% compared to the theoretical density of pristine YAG. The density plot clearly reveals that the materials fabricated at 500°C and 700°C, has substantial fraction of materials in amorphous form. SEM images observed at low magnification clearly reveals any form of porosity present in the material and clearly shows the ability to consolidate them very well under the chosen experimental conditions (**Figure 3c**). However, when the materials surface was observed under higher magnification, some pores with a diameter on the order of ~100nm are seen (**Figure 3d**). We attribute the ability to retain the amorphous structure while achieving the densification of particles into bulk form to the optimized processing conditions, especially to the low calcination and also the densification temperature. In addition, with achieving the amorphous material in bulk, dense form, we were also able to achieve partially crystalline and fully crystalline 3 at.% Ce:YAG.

**C. Phase and Microstructural Analysis of Ce:Al$_2$O$_3$**

In **Figure 4a** the effect of CAPAD processing temperature on the relative density of 0.5 at.% Ce:Al$_2$O$_3$ ceramics is plotted. We observe the classic sigmoidal behavior with the densification onset temperature of ~1150°C and the relative density nears the theoretical limit (>99%) at processing temperatures above 1250°C with a 5 min isotherm under 105 MPa. The transparency of the sample produced at 1250°C can be appreciated in the photo included as an inset in **Figure 4a**. Under these processing conditions, the Ce:Al$_2$O$_3$ specimen has an average grain size of ~300nm (**Figure 4b**) and is phase pure as determined by XRD of the bulk ceramics (**Figure 4c**), when compared to the Al$_2$O$_3$ diffraction pattern standard (ICSD# 63647). The transparency which can be appreciated in **Figure 4d**, where the real in-line transmission is plotted as a function of wavelength from the UV through the mid IR. In addition, we present an optical micrograph of a polished transparent ceramic placed atop black text as an inset that qualitatively depicts the sample high transparency of the bulk ceramic. The transparency is attributed not only to the elimination of residual porosity, but to the phase purity and fine average grain size, as scattering due to refractive index mismatch at pores and grain boundaries are minimized. These results are in agreement with our previous reports on the production of laser grade ceramic ruby[32] and Tb:$Al_2O_3$[33] and Nd:$Al_2O_3$[25], all of which were produced from similar powders and CAPAD processing conditions. However, in contrast to our observations in the Nd:$Al_2O_3$ system, we do not observe a systemic stretching of the matrix lattice in the Ce:$Al_2O_3$ system and this may merit further investigation and are currently outside of the scope of this manuscript.

In addition to influencing the grain boundary scattering, the small average grain sizes permit the over-equilibrium incorporation of Ce within the $Al_2O_3$ matrix, which leads to optical activity from the Ce-ion, with the latter being discussed in **Section II-D**. The small grain size results in an increased grain boundary volume, resulting in space for the dopants to distribute, permitting the matrix to accommodate higher dopant concentrations with an increased average distance between Ce-ions, avoiding concentration quenching effects. For alumina, the critical grain size needed to fully accommodate dopant concentrations required for PL applications ($c_{vol}$~$10^{20}$ *ions/$cm^3$*), with an optimal inter-ionic distance to avoid PL quenching is ~25nm. Currently, the production of such microstructures is not experimentally possible, with the final grain size at full density being a limiting factor. For a detailed discussion about the grain boundary volume and volumetric dopant concentrations please refer to our recent communication.[25] As the actual grain size of Ce:$Al_2O_3$ is ~300nm, the accommodation of optically active dopants along grain boundaries is not possible in these materials, indicating that the microstructure cannot fully explain the presence of PL active Ce within the bulk matrix. Fortunately, this requirement is somewhat relaxed as the applied CAPAD processing conditions are kinetically favorable (high heating and cooling rates) to result in a quenched microstructure with rare-earth present within grain interiors[25][28][33]. In this case, the dopant concentration is ~500 times above the solubility limit, i.e. at volumetric concentrations similar to a common Ce dopant concentration in Ce:YAG phosphors[6][25].

## D. PL Properties of Ce:YAG and Ce:Al$_2$O$_3$

**Figure 5a** plots the PL excitation and emission scans for 3 at.% Ce:YAG powders as a function of calcination temperature over which the powders transition from fully amorphous to fully crystalline. This dopant concentration is chosen as the PL intensity is sufficient to produce adequate down conversion while avoiding detrimental concentration quenching. These effects have been investigated by a number of independent researchers, including our previous work on bulk Ce:YAG ceramics, which showed that an optimal Ce dopant concentration range is on the order of 3-4at.%[23][34].

The excitation scans show the classic absorption bands associated with the 5d←4f transitions, with the 4f-5d$_2$ centered at 330nm and the 4f-5d$_1$ centered at 465nm, consistent with our previous work in solid-state reacted Ce:YAG[28]. Additionally, the PL measurements show that the absolute intensity of the excitation and emission spectra monotonically increase with the degree of crystallization, with no optical activity being present in the fully amorphous powders that were calcined at 800°C. These results indicate that it is the interaction of the Ce atom with the ordered structure of the YAG lattice that results in optical activity and are consistent with previous studies of Ce:YAG powders produced through other routes[35].

Pumping the Ce:YAG with wavelengths accessing the absorption bands, yields the broadband luminescence from ~500nm-650nm with its peak at approximately 530nm, which arises from the 5d band splitting into 5 levels and relaxing to the doubly degenerate 4f ground state[36]. However, since powders cannot be readily characterized at cryogenic temperatures, we were not able to resolve the splitting of the 4f ground state. The emission scans were taken with pumping at 465nm, the most efficient pumping wavelength and the typical pumping wavelength that is used to produce white light from Ce:YAG via color mixing. Since Ce:YAG does not have emission in the blue portion of the spectrum, (<500nm), engineers typically adjust the thickness of the phosphor capping layer to permit some "bleed-through" of the blue pumping source, which tend to have a rather narrow FWHM of ~5-10nm. In order to improve the color rendering, a contribution from an additional phosphor capping layer (e.g. Ce:$Al_2O_3$, **Figure 1b-c,** and discussion below) with a broader FWHM in the blue portion of the spectrum a larger can be incorporated within a SSWL device. In the Ce:YAG/Ce:$Al_2O_3$ case, the pumping can be done with a single UV LED since the absorption bands overlap.

We also report a similar trend in bulk Ce:YAG materials that are fully dense, yet possess differing degrees of crystallization as described in **Section II-B**. **Figure 5b** shows the room temperature excitation and emission PL spectra for three samples, a fully amorphous sample, a partially crystalline/amorphous sample, and a fully crystalline sample. These results reveal the high dependence of the optical activity to the structure of the host; the ceramic with a nearly amorphous structure shows no optical luminescence activity, while those samples with crystalline Ce:YAG present show optical response. Similar to the powder samples, the absolute intensity increases with the increased volumetric proportion of crystallinity phase present. The line shape and intensity of the excitation and emission scans of the fully crystalline material is also consistent with previous reports on the transitions of $Ce^{3+}$ in YAG[28][36]–[39].

In addition to room temperature PL spectroscopy, the bulk nature of the sample permitted spectroscopic characterization across the temperature range from 298K to 4K, which is plotted in **Figure 5c**. By subjecting the samples to low-temperatures the thermal effects of the host on the Ce-ions can be mitigated and the line broadening and splitting of the 4f ground states can be observed. Following the methodology of Nair[39], we were able to de-convolute the excitation and emission spectra for the bulk sample and confirm the energy gaps for the 4f-5$d_2$ and 4f-5$d_1$ transitions, as well as the splitting of the 4f states. These measurements are in line with the energy diagram for Ce:YAG, reconstructed herein in **Figure 7a**, with data from the previous reports by Tomiki[36], Hamilton[37], Zych[38], including the location of the zero phonon-line, i.e. the intersection of the excitation and emission spectra, which is measured here at room temperature to be ~515nm and at 4K to be blue-shifted to ~500nm. Since the lower 4f state is nearly depopulated at room temperature, it is likely that at cryogenic temperatures the splitting is thermally mitigated, resulting in a lesser contribution towards the red portion of the spectrum. Here however, the spectroscopic techniques do not permit the observation of the higher energy 5d states, which require more sophisticated excited state-absorption and photoconductivity measurement methods[37][38]. Our methods do permit the measurement of the excitation and emission spectra to temperatures below 60K, which were omitted in the work of Zych because the sensitivity of the detector was unable to resolve any PL signature at lower temperatures.

In addition, we observe that the intensity of the PL spectra decreases at low-temperatures, which is not the typical behavior for rare-earths, as reduction in temperature usually results in line-thinning and accompanied by an increased PL intensity[19][38]. The atypical behavior has been ascribed to the interaction of lattice traps with the electrons that are brought to the higher energy 5d energy states of the Ce dopant that overlap with the conduction band as shown in **Figure 7a**. At low temperatures the traps have more time to interact with the free electrons in the conduction band, offering a non-radiative pathway to relaxation, resulting in decreased PL intensity[38]. This effect is reduced with increased temperature, thereby reducing non-radiative decays which result in an increase in the PL intensity.

The partially crystalline sample however, does show interesting behavior in the line-shape and the position of the zero-phonon line. **Figure 5d** plots the PL spectra of the two Ce:YAG samples that have on a normalized scale, which allows easier comparison of the line-shape characteristics. The excitation scan of the partially crystalline samples reveals the absence of the 4f-5d$_2$ band centered at 330nm, as well as a narrowed FWHM of the 4f-5d$_1$ band centered at 465nm, from ~93nm to ~58nm. This behavior indicates that although XRD does not show any shifts in the peak position of the YAG peaks, that the symmetry of the crystalline phase has not yet fully approached the D2 symmetry which cause the 5d states to split into five components[36]. Consequently, the resultant emission is also narrowed from ~105nm to 67nm as the higher energy 5d states are not present to present optical absorption and branch to the 4f ground state upon relaxation. The lack of emission towards the red suggests that the ground-state has not fully split and the shift in the position of the zero-phonon line from ~515nm to ~497nm is evidence that the energy difference between the $5d_1$ and $4f_1$ is smaller than in fully crystalline Ce:YAG by as much as 0.1eV, which is consistent with a crystal field interaction similar to what is observed in the fully crystalline sample at cryogenic temperatures.

As a side-note, in this sample, a small emission peak corresponding to the R-lines for ruby ($Cr:Al_2O_3$) at 693nm is clearly discerned. This is likely as Cr is present as a minor impurity within the aluminum nitrate salts used for synthesis. Although the XRD does not show the presence of the secondary phase, the PL characterization is thus sensitive enough to discern intermediary stages of crystallization to YAG. The ruby R-line luminescence is not present in the fully crystalline ceramics as alumina is no longer present at this stage since it is consumed to produce YAG. This is a plausible scenario, as the phase detention limits is on the order of 1-5%, which extremely low (ppm) amounts of active ions can be detected using modern photomultiplication equipment and techniques. These measurements show the benefit of corroborating phase identification measurements with supporting characterization such as PL instead of fully relying on standard XRD techniques for phase identification throughout processing as is often done.

In a similar manner we have characterized the PL excitation and emission spectra for $Ce:Al_2O_3$ from room to cryogenic temperatures. We observe that the PL intensity of $Ce:Al_2O_3$ increases at low temperatures (**Figure 6a**), which is in stark contrast to the classic phosphor down-convertors like $Ce^{3+}$:YAG. **Figure 6b** plots the peak intensity over the temperature range for the normalized excitation and emission scans of Ce:Al2O3. Comparison of **Figures 6a** and **6b** allows one to clearly see the differences in the trends between the two material systems. The peak PL intensity for $Ce:Al_2O_3$ increases by a factor of ~1.8 while it decreases to about 30% of the room temperature value for Ce:YAG. Using the same fitting approach described above on the spectra collected at 4K, we were able to determine the location of the $5d_1$-$5d_4$ states (**Figure 6c**) and were also able to resolve the crystal-field splitting of the ground state ($4f_1$ and $4f_2$, **Figure 6d**) to be ~0.36eV, slightly larger than in Ce:YAG. We assume that the $5d_5$ state requires higher energy (lower wavelengths) to access than UV-Visible pumping as is the case for Ce:YAG.

The spectra also show evidence of line thinning, particularly in the excitation spectra, as the $5d_4$ at ~300nm is favored at lower temperatures, while some of the structure pertaining to the lower 5d states is observable at room temperature. The behavior of Ce:$Al_2O_3$ is consistent with the low temperature optical behavior of other rare-earths doped into oxides, such as Nd[41] and Er[42] doped YAG, that exhibit optical 4f to 4f transitions that are shielded from crystal field interactions by the outer 5d shell. The low temperature behavior of Ce:$Al_2O_3$ indicates that the position of the 5d←4f optical transitions are located below the conduction band, which allows us to propose the simplified energy band diagram for Ce in $Al_2O_3$ in **Figure 7b**. These measurements do not allow for the determination of the exact position of the ground state in relation to the conduction and valence bands. Precise locations could be determined using low-temperature transmission measurements of these materials, which is currently unavailable in our experimental set-up. As such we must leave the precise location of the 4f ground state in relation to the valence band as an open question for a subsequent communication.

### III. Summary

Bulk 0.5 at.% Ce:$Al_2O_3$ ceramic phosphors were fabricated with a simultaneous solid-state reactive densification route via CAPAD. The highly doped Ce:$Al_2O_3$ phosphors have a broadband emission in the blue/green portion of the optical spectrum from 400-600nm when pumped with UV light. In addition, Ce:YAG powders and bulk Ce:YAG ceramics were fabricated with controlled crystallinity. The PL spectra were characterized and compared to Ce:$Al_2O_3$ from 298K to 4K, revealing that the different crystal fields result in different absorption and emission. Using these characterization tools we are able to propose the energy differences between the 5d and 4f states of Ce in $Al_2O_3$ and the increased PL intensity over the same temperature range indicates that the observed 5d states likely do not overlap with the conduction band as has been observed in

Ce:YAG. Instead Ce:Al$_2$O$_3$ behaves similar to other rare-earths, with an observed line thinning an increase in spectral intensity at lower temperatures, indicative of quenching of thermally induced broadening. The over-equilibrium doped phosphors bring wide-band PL from the Ce$^{3+}$ active ion transitions to a matrix with a higher thermal conductivity and thermal shock resistance. Ce:Al$_2$O$_3$ can be used with other broadband phosphors such as Ce:YAG as an active encapsulant or as an additional capping layer to produce LD pumped high-brightness/high-power SSWL devices with improved thermomechanical performance that don't rely on LED pumping which suffers from efficiency droop.

## IV. Methodology

### A. Synthesis, Processing and Densification of Ce:Al$_2$O$_3$

Prior to reaction-densification, ultra-high purity α-Al$_2$O$_3$ sub-micron powder (99.99% purity, Taimei Chemical, Japan) and CeO$_2$ nano-powder (99.97% purity, Sigma-Aldrich, USA) were weighed and mixed to achieve a Ce:Al ratio of 0.5%. After mixing in a mortar, the powders were suspended in ultra-high purity water (UHP, 99.99% purity) and tumble milled at ~50RPM using 3mm diameter, α-Al$_2$O$_3$ spherical media (99.99% purity, Taimei Chemical, Japan). The mixed powders were sieved and dried at 70°C under 30mmHg vacuum. Following vacuum drying, 3.000 ± 0.0001g of the powders were suspended in 30mL of UHP water and planetary ball milled at 150 RPM for 3 and 6 hrs using 30g of 3mm diameter media. The powders were sieved and dried in air at 120°C for 12 hrs and kept dry until consolidation.

The bulk ceramic Ce:Al$_2$O$_3$ phosphors were produced using an all solid-state, one-step reaction-densification route using Current Activated Pressure Assisted Densification (CAPAD)[23][42]. 1.500g ± 0.0001g of powder was secured within graphite tooling, to produce fully dense bulk disk-shaped specimens with a 19mm diameter. The ceramic specimens were processed at various temperatures with a heating rate of ~300°C/min, and a 5min isothermal treatment. In conjunction with the thermal cycle, the powders were uniaxially pressurized to 105

MPa with a 35.33 MPa/min pressure ramp. The ultimate uniaxial pressure was released at the end of the isothermal treatment.

**B. Synthesis, Processing and Densification of Ce:YAG**

The 3 at.% Ce:YAG ($Y_{2.97}Ce_{0.03}Al_5O_{12}$) powders used in this study were synthesized through a solution-polymerization based route, organic/inorganic steric entrapment method[29]–[31]. High purity Yttrium(III) nitrate hexahydrate, Aluminum Nitrate nonahydrate and Cerium(III) nitrate hexahydrate (Alfa Aesar, Ward Hill, MA, USA) were the precursor sources of the cations. The single-pot synthesis route began with preparing polymeric solution containing a 5 wt.% of 80% hydrolyzed polyvinyl alcohol (PVA, Sigma-Aldrich, St. Louis, MO, USA) dissolved in deionized water by stirring for 24 h at room temperature. Stoichiometric amounts of $Y(NO_3)_3 \cdot 6H_2O$, $Al(NO_3)_3 \cdot 9H_2O$ and $Ce(NO_3)_3 \cdot 6H_2O$ were mixed together, following the stoichiometry $Y_{2.97}Ce_{0.03}Al_5O_{12}$, and stirred with PVA solution. This solution was stirred for an additional 12 hr. The PVA fraction in solution with the nitrates was chosen so that there were 4 times more positively charged valences from the cations than negatively charged functional end groups of the organics (in the case of PVA, −OH groups)[30]. This strategy ensures that there are more cations in the solution than the hydroxyl functional groups of the polymer, with which they could chemically bond. The precursor solutions were then heated on a hot plate with continuous stirring (350°C, 300 RPM) until the water evaporates, resulting in a thick yellow gel. The aerated gel formed was then vacuum dried at 70°C under -850 millibar for 24 hr. The vacuum drying process resulted in a yellowish dried crisp foam which was first ground using an agate mortar and pestle. The ground powders were heat treated at various temperature for 30 minutes to investigate the phase evolution behavior. To fabricate bulk specimens, as synthesized powders were calcined at 700°C for 30 minutes which results in pale yellowish powders. The calcined powders were then planetary ball milled (equipment details and also 300 RPM) for 6hr using 10mm silicon nitride ($Si_3N_4$) milling media with de-ionized water to reduce the particle size and to increase the specific surface area. The milling media were then drained and the solution was centrifuged at 3300 RPM for 10 min, then dried at ~ 70°C for 24 hr using vacuum drying oven under -850 millibar. The dried granules were ground in a mortar and stored dry until CAPAD processing.

These amorphous powders were consolidated using low-temperature and high-pressures with custom built CAPAD apparatus to obtain materials with varying crystallinity. The high-heating rates (~300°C/min) associated with CAPAD allow control over the total processing times, which in conjunction with the high-applied pressures (~500 MPa), allow the fabrication of fully dense ceramics with control in the ultimate crystallinity. 0.25 g of powders were packed into 10 mm diameter WC-Co die, initially prepressed under 500 MPa for 2 minutes and the pressure was completely removed before starting the experiment. Pressure was increased to ~70 MPa/min while the die setup was heated at the rate of ~300°C/min. A fully dense fully crystalline Ce:YAG sample was produced, also via CAPAD, but under conventional processing conditions of 1400°C and 100MPa within graphite tooling, with all other parameters unchanged. This was done because the WC-Co tungsten carbide tooling is not suitable for operation above ~1000°C, which yielded only partially crystalline ceramics (not presented herein).

### C. Material Characterization

Differential thermal analysis (DTA) and thermogravimetric analyses (TGA) (SDT Q600-TA Instruments, New Castle, DE) at 30°C/min were performed on the as-synthesized powders to study the precursor-to-ceramic powder conversion and phase formation. X-ray diffraction (XRD) using Cu Kα1 ($\lambda$=1.54058 Å) radiation was implemented on a Phillips X'Pert Diffractometer (Model DY1145), in point source mode, with a 45kV potential, 40mA current, a 0.002 step-size, and 2s integration time, from 2Θ=15-60°. The collected XRD spectra were compared to published standards for α-$Al_2O_3$ (ICSD#: 63647) and YAG (ICSD#: 23848).

The bulk density of the bulk specimens was measured by the Archimedes' method using deionized water at room temperature. A low-speed saw was employed to section the densified specimen using a diamond tipped saw, and cross-sectioned regions were grounded using diamond impregnated grit disks for microstructural investigations. Scanning electron microscopic (SEM, Zeiss Sigma 500) analyses was performed on the surfaces of the specimens that were polished with diamond lapping (grit~0.1µm), as well as the powders and no conductive layer was applied prior to observation.

**D. Room Temperature and Cryogenic Photoluminescence Characterization**

Photoluminescence (PL) excitation and emission spectra were measured on a custom built spectrofluorometer based on Horiba-PTI model 3558, using a tungsten deuterium lamp with monochromators as an excitation source. The spectroflourometer was coupled to an Advanced Research Systems closed-cycle helium cryostat (ARS-2HW) using a conductive cold finger within a windowed housing that enabled measurements from 4K to 295K. All measurements were performed at 45° angle of incidence (AOI) on polished bulk ceramics and the PL spectra were collected from the specimen front face. Excitation spectra were collected between $\lambda = 250$nm and $\lambda = 420$nm, while monitoring the peak emission intensity near $\lambda = 465$nm. Emission spectra were collected between $\lambda = 350$nm and $\lambda = 610$nm, while monitoring the peak emission intensity near $\lambda = 330$nm. In all cases, the excitation and emission monochromators were adjusted to achieve a spectral bandwidth of 1nm. The step-size was 1nm, while the intensity was integrated for 1s. Three excitation and emission spectra were collected and averaged.

**Acknowledgements:** We gratefully acknowledge the funding of this work by a multidisciplinary research initiative (MRI) from the High Energy Lasers—Joint Technology Office (HEL-JTO) and the Office of Naval Research (ONR). Some of the characterization work was enabled by ACIF grant NSF CHE-9974924.

**Figures and figure Captions**

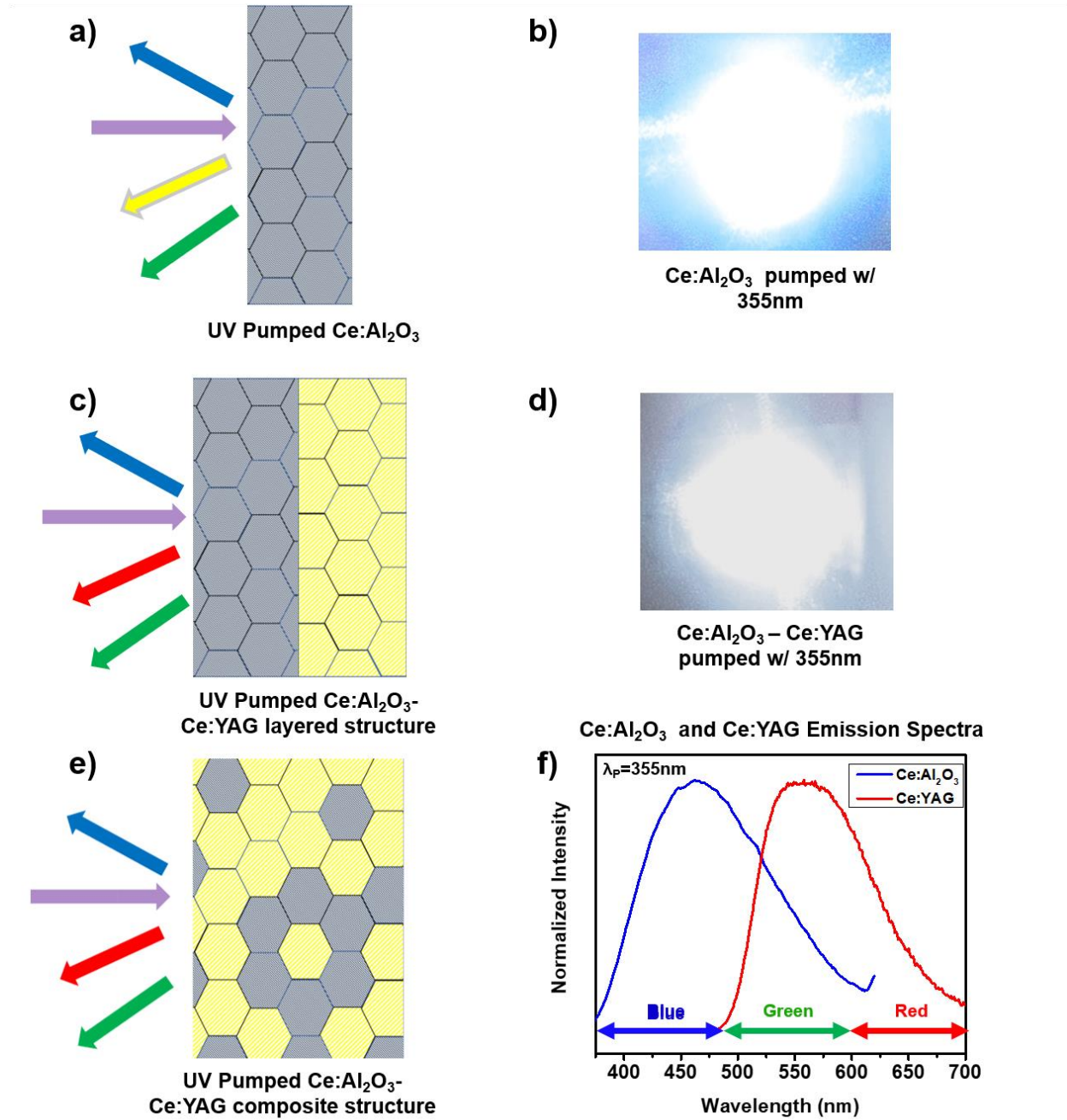

Figure 1 a) Conceptual schematic of Ce:Al₂O₃ as a stand-alone UV-pumped broadband blue-green phosphor, b) Picture of bulk Ce:Al₂O₃ emission when pumped with 355nm laser light, c) Conceptual schematic of Ce:Al₂O₃/Ce:YAG layered structure for UV laser pumped SSWL, d) Picture of the resulting emission from the Ce:Al₂O₃/Ce:YAG layered structure when pumped with

355nm laser light, e) Conceptual schematic of Ce:Al$_2$O$_3$-Ce:YAG composite based on the design of Denault[9], Cozzan[8] and Wang[11], and f) Emission spectra of Ce:Al$_2$O$_3$ and Ce:YAG, showing the overlap in emission that would fully encompass the blue, green, and extend into the red.

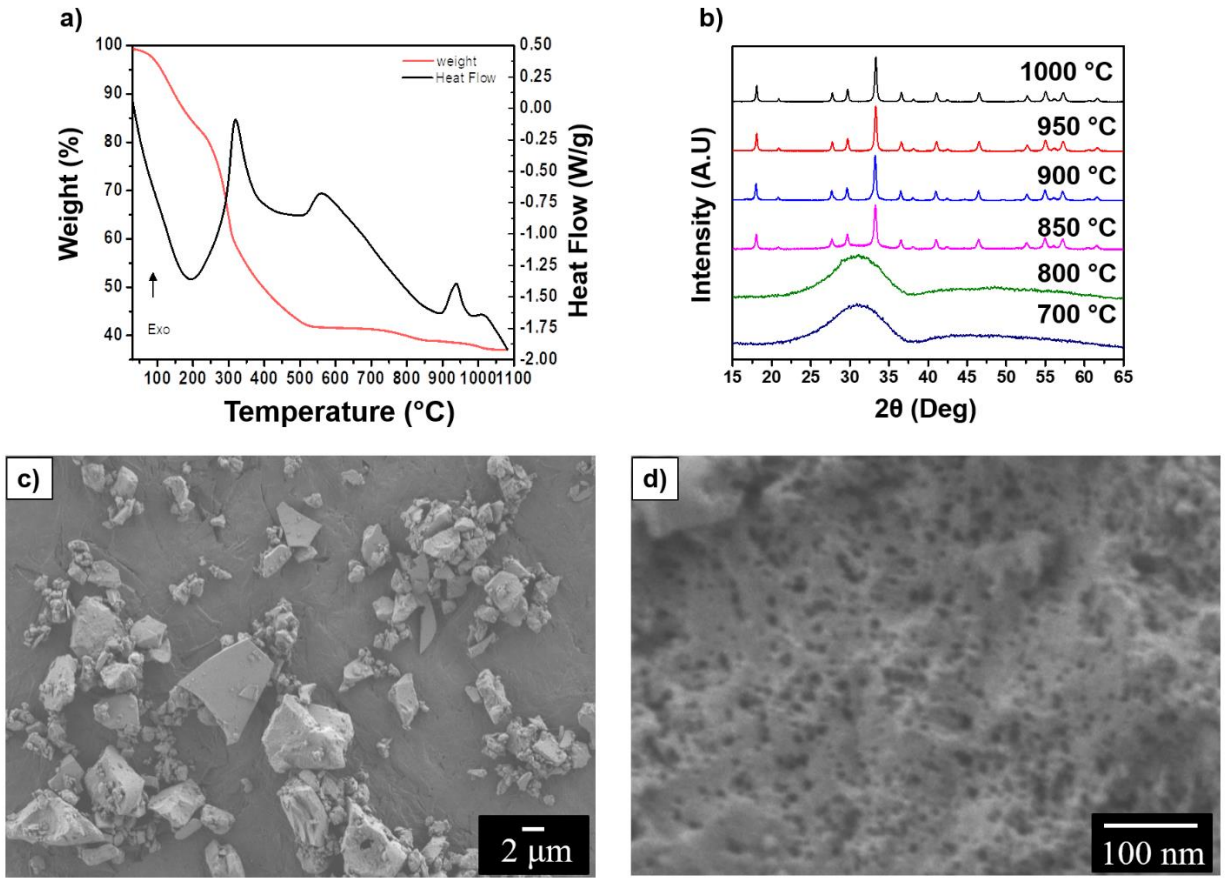

Figure 2 a) DTA and TGA of as-synthesized 3 at.% Ce:YAG powders prior to calcination, b) XRD of Ce:YAG powders vs. annealing temperature, c) Low-magnification SEM of Ce:YAG powders annealed at 700°C for 30min, and d) High-magnification SEM of Ce:YAG powders annealed at 700°C for 30min, showing the typical morphology of these as synthesized powders.

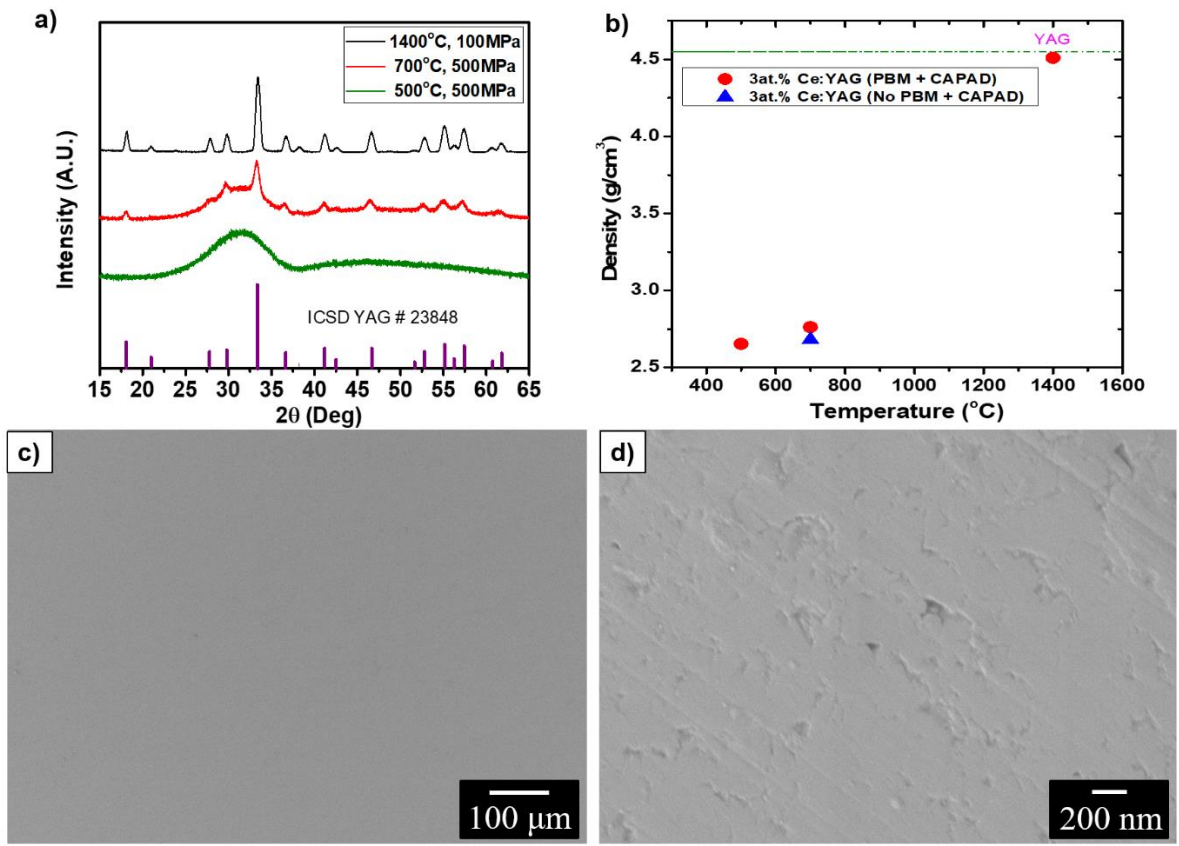

Figure 3 a) XRD of bulk Ce:YAG ceramics for different CAPAD processing conditions, b) Plot of the Relative Density vs. CAPAD processing temperature for Ce:YAG, c) Low magnification SEM of the bulk Ce:YAG produced at 500°C and 500MPa, showing lack of porosity even though the bulk-density is significantly lower than that of fully crystalline Ce:YAG, and d) High magnification of the bulk Ce:YAG produced at 500°C and 500MPa, showing the presence of a small amount of pores of ~100nm diameter within the bulk ceramic.

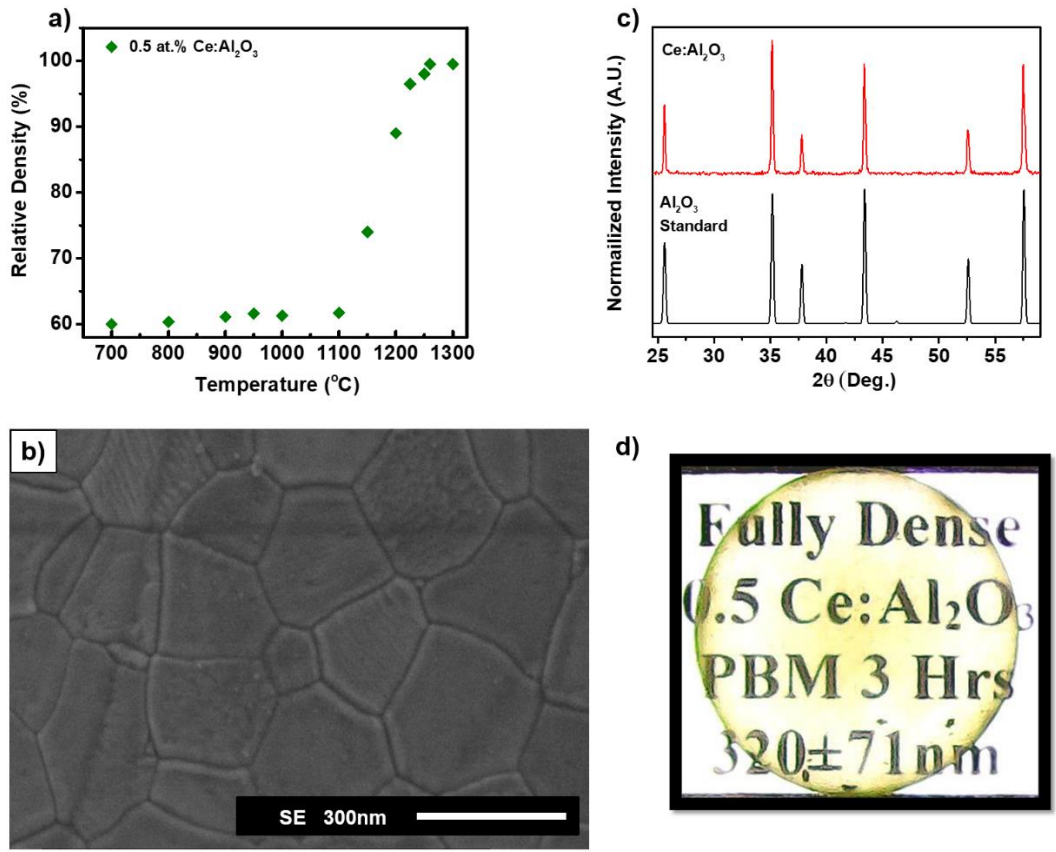

Figure 4 a) Plot of the Relative Density vs. CAPAD processing temperature for Ce:$Al_2O_3$, b) SEM of polished and thermally etched bulk Ce:$Al_2O_3$ processed at 1250 °C, c) XRD of bulk Ce:$Al_2O_3$ compared to the diffraction pattern for pure α-$Al_2O_3$ (ICSD#: 63647), and d) Picture of a Ce:$Al_2O_3$ ceramic taken atop a text, showing the sample transparency.

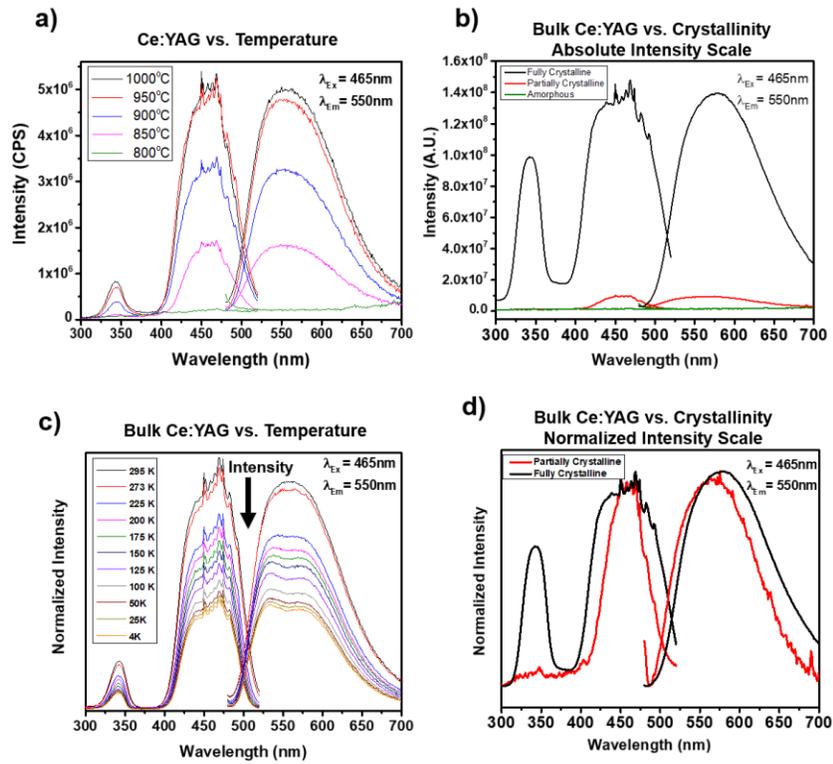

Figure 5 a) PL excitation and emission spectra of 3at.% Ce:YAG powders annealed at different temperatures, b) Absolute Intensity PL excitation and emission spectra of bulk amorphous and partially and fully crystalline 3at.% Ce:YAG ceramics, c) PL excitation and emission spectra of bulk crystalline Ce:YAG from room to cryogenic temperatures, d) Normalized Intensity PL excitation and emission spectra of bulk, partially and fully crystalline 3at.% Ce:YAG

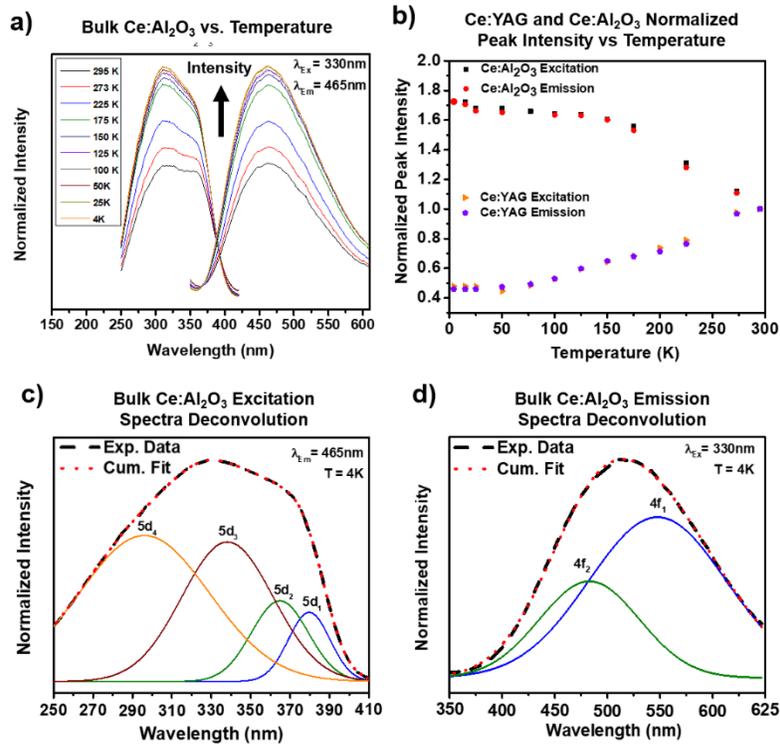

Figure 6 a) PL excitation and emission spectra of bulk crystalline Ce:Al$_2$O$_3$ from room to cryogenic temperatures, b) Plot of the Peak Normalized Intensity from room to cryogenic temperatures for Ce:YAG and Ce:Al$_2$O$_3$ c) PL excitation spectra of Ce:Al$_2$O$_3$ collected at 4K, with a cumulative fit of the 5d-states superimposed on the experimental data, and d) PL emission spectra of Ce:Al$_2$O$_3$ collected at 4K, with a cumulative fit of the 5d-states superimposed on the experimental data.

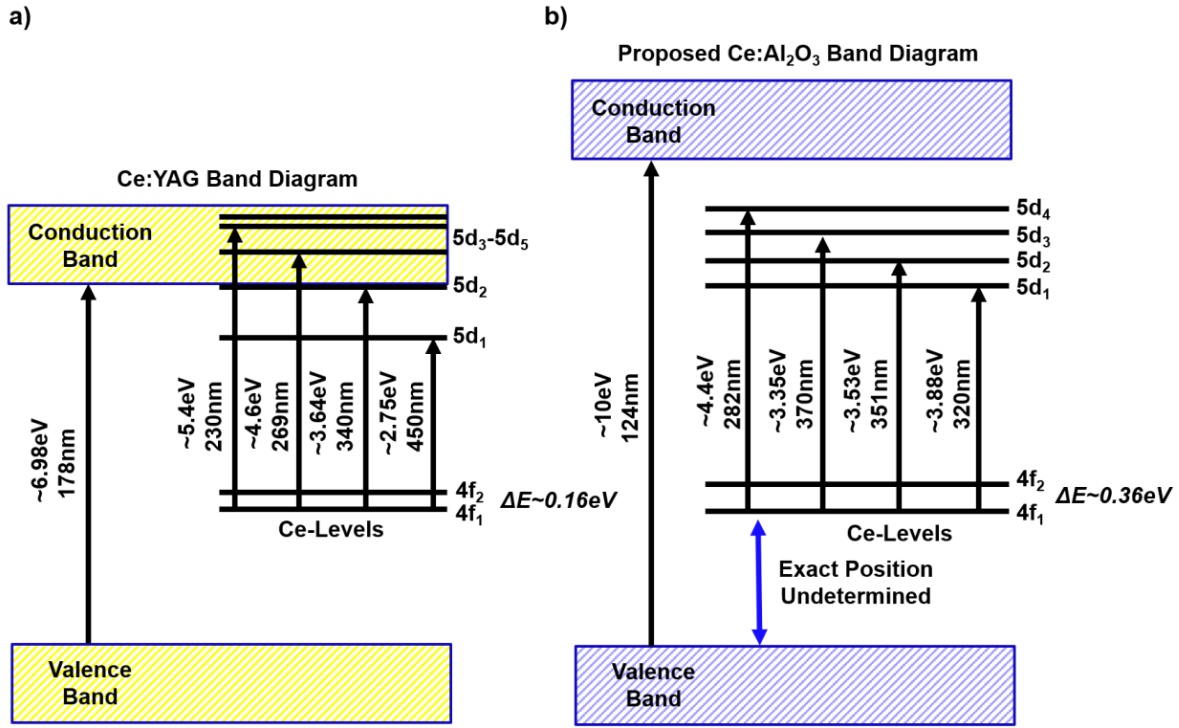

Figure 7 a) Ce:YAG Band Diagram as previously suggested using data from Tomiki[36], Hamilton[37], and Zych[38] and b) Proposed simplified Band Diagram for Ce:Al$_2$O$_3$.